\documentclass[aps,pra,twocolumn,groupedaddress,showpacs]{revtex4}

\usepackage{graphics}

\begin{document}


\title{Multiscale quantum-defect theory for two interacting atoms in a symmetric harmonic trap}


\author{Yujun Chen}
\author{Bo Gao}
\email[Email:]{bgao@physics.utoledo.edu}
\homepage[Homepage:]{http://bgaowww.physics.utoledo.edu}
\affiliation{Department of Physics and Astronomy,
	University of Toledo, MS 111,
	Toledo, Ohio 43606}


\date{\today}

\begin{abstract}

We present a multiscale quantum-defect theory (QDT) for two identical 
atoms in a symmetric harmonic trap
that combines the quantum-defect theory for the van der Waals 
interaction [B. Gao, Phys. Rev. A \textbf{64}, 010701(R) (2001)]
at short distances with a quantum-defect theory for the harmonic trapping 
potential at large distances. The theory provides a systematic understanding 
of two atoms in a trap, from deeply bound molecular states and states of
different partial waves, to highly excited trap states. It shows, e.g., that
a strong $p$ wave pairing can lead to a lower energy state around the 
threshold than a $s$ wave pairing.

\end{abstract}


\maketitle

\section{Introduction}

Two interacting particles under confinement, described generally 
by a Hamiltonian
\begin{equation}
H = -\frac{\hbar^2}{2m_1}\nabla_1^2-\frac{\hbar^2}{2m_2}\nabla_2^2
	+V_1(\mathbf{r}_1)+V_2(\mathbf{r}_2)+v(|\mathbf{r}_1-\mathbf{r}_2|) \;,
\end{equation}
where $V_1$ and $V_2$ are the confining potentials and
$v(r)$ is the interaction between particles,
represents a fundamental class of problems in physics.
One famous example is the helium-atom problem that 
has played an important role in our understanding of
electron correlation in atomic physics (see, e.g., Ref.~\cite{fan86}).
Similarly, the problem of two atoms in a harmonic trap, 
which has attracted considerable recent attention 
(see, e.g., Refs.~\cite{bus98,ols98,tie00,blu02a,blo02,bol02,ber03,
bol03,gra04,mor05,sto06,idz06}),
is the key to our understanding of atomic correlation in
a trapped many-atom quantum system. 
Such correlation differs qualitatively from the electron
correlation because atoms attract each other at large distances 
and can form bound states.

Existing theories of two atoms in a trap have 
relied mostly upon the pseudopotential model of atomic 
interaction \cite{hua57}, and its generalizations \cite{bol02}.
While such models can work well in describing
how the trap states, especially the lowest few, are affected
by atomic interaction, they generally fail in
describing how a molecular state is affected by trapping, 
with the only exception being the least bound molecular state with
a very large scattering length.
Furthermore, such theories do not adequately address nonzero partial
waves, for which naive generalizations of the ``shape-independent''
approximation \cite{bus98}, using, e.g., the effect range 
theory (ERT) \cite{bla49}, would generally lead to incorrect results.

We present here a multiscale (2 length scales, to be exact) 
quantum-defect theory (QDT) for two identical atoms in
a symmetric harmonic trap.
It is a completely general theory that works for different
partial waves, and from deeply bound molecular states to 
highly excited trap states.
In Sec.~\ref{sec:qdt}, we expand our tool box of QDT for
different long range potentials \cite{gre79,gre82,fan86,gao98a,gao99a} 
by presenting a QDT for a symmetric harmonic potential. 
It is independently useful beyond the scope of two atoms
in a trap. For example, it may be used to treat
two nucleons outside of a closed shell \cite{fet71}.

In Sec.~\ref{sec:tasht}, this theory is combined with the 
angular-momentum-insensitive quantum-defect theory (AQDT) for
the van der Waals interaction \cite{gao01} to 
formulate a two-scale QDT that provides a systematic understanding
of two identical atoms in a symmetric harmonic trap.
Results and discussions are presented in Sec.~\ref{sec:results},
including a discussion of the limitations of ``shape-independent'' 
approximations, and a universal spectrum for two atoms in a trap 
at the van der Waals length scale \cite{gao01,gao04a,gao05b,kha06}
that shows, e.g., that a strong $p$ wave pairing can lead to a lower 
energy state around the threshold than a $s$ wave pairing.
We will also show that two atoms
in a trap has long-range correlation that becomes important for
large scattering lengths, a result that has proven to be the key
for generalizing the variational Monte Carlo (VMC)
studies of few atoms in a trap to the regime of strong 
coupling \cite{kha06}.
Conclusions are given in Sec.~\ref{sec:conclusions}.

\section{Quantum-defect theory for a symmetric harmonic potential
	\label{sec:qdt}}

The goal of a QDT for a symmetric harmonic potential is to
provide a systematic understanding to a class of problems
described by the radial Schr\"{o}dinger equation
\begin{equation}
\left[-\frac{\hbar^2}{2\mu}\frac{d^2}{dr^2} 
	+ \frac{\hbar^2 l(l+1)}{2\mu r^2}
	+ V(r) - \epsilon \right]
	u_{\epsilon l}(r) = 0 \;,
\label{eq:rsch}
\end{equation}
with
\begin{equation}
V(r)\stackrel{r\rightarrow\infty}{\longrightarrow} 
	\frac{1}{2}\mu\omega^2r^2 \;.
\label{eq:hopotdef}	
\end{equation}
Unlike the standard textbook solution which requires that
$V(r)=\frac{1}{2}\mu\omega^2r^2$ for all $r$, a case that we shall
refer to as the ``pure'' harmonic oscillator,
the QDT formulation is applicable to any $V(r)$ that is 
asymptotically a harmonic oscillator,
but may differ from it at short distances in an arbitrary fashion.

As in any QDT formulation \cite{gre79,gre82,fan86}, we start by defining 
a pair of reference functions that are two linearly independent solutions 
for a symmetric harmonic potential
\begin{equation}
\left[-\frac{\hbar^2}{2\mu}\frac{d^2}{dr^2} 
	+ \frac{\hbar^2 l(l+1)}{2\mu r^2}
	+\frac{1}{2}\mu\omega^2r^2- \epsilon \right]
	v(r) = 0 \;.
\label{eq:horsch}
\end{equation}
The solutions can be easily found \cite{abr64}, and we will take 
\begin{eqnarray}
f^{(ho)}_{e l}(x) &=& 
	x^{l+1}e^{-x^2/2}M(b,c,x^2)  \;, 
\label{eq:fho}\\
g^{(ho)}_{e l}(x) &=& -\frac{2}{(2l+1)\pi}
	x^{-l}e^{-x^2/2} \nonumber\\ 
	& &\times M(1+b-c,2-c,x^2) \;.
\label{eq:gho}
\end{eqnarray}
Here $e = \epsilon/\hbar\omega$ is a scaled energy.
$x =r/\beta_{ho}$ is a radius scaled by 
$\beta_{ho}=(\hbar/\mu\omega)^{1/2}$, which is a length scale
associated with the harmonic potential. 
$M$ is the confluent hypergeometric function \cite{abr64},
$b = (l+3/2-e)/2$, and $c=l+3/2$.
In this definition,
$f^{(ho)}_{e l}$ is regular at the origin,
and $g^{(ho)}_{e l}$ is irregular. They will be
called the regular solution and the irregular solution, respectively.
They are also chosen such that their Wronskian is given by
\begin{equation}
W(f^{(ho)}_{\epsilon_s l},g^{(ho)}_{\epsilon_s l})
	\equiv f^{(ho)}_{\epsilon_s l}\frac{d g^{(ho)}_{\epsilon_s l}}{dx}
	-g^{(ho)}_{\epsilon_s l}\frac{d f^{(ho)}_{\epsilon_s l}}{dx}
	=\frac{2}{\pi} \;.
\end{equation}

With this definition of reference functions, the wave function
$u_{\epsilon l}$ for any potential that is asymptotically a harmonic oscillator
can be written, at sufficiently large distances, as
\begin{equation}
u_{\epsilon l}(r) = A_{\epsilon l}[f^{(ho)}_{e l}(x) 
	- K^{(ho)}(\epsilon,l) g^{(ho)}_{e l}(x)]\;.
\label{eq:wfn}
\end{equation}
This defines the $K$ matrix, $K^{(ho)}$, for a symmetric
harmonic potential, with its value being generally determined by
matching Eq.~(\ref{eq:wfn}) to the short-range solution.

\begin{figure}
\scalebox{0.4}{\includegraphics{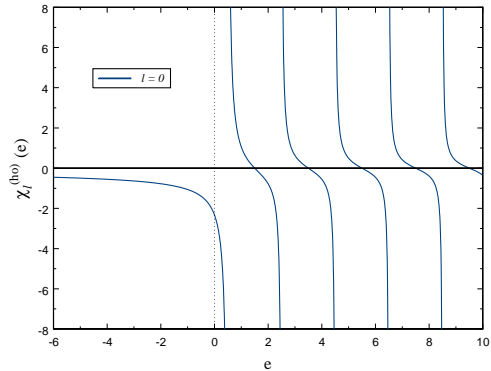}}
\caption{The $\chi^{(ho)}_l(e)$ function for the $s$ wave.
The energy spectrum for any potential that is asymptotically a 
symmetric harmonic
oscillator is given by the crossing points of this function with
a short-range $K$ matrix $K^{(ho)}(\epsilon, l)$ defined by
Eq.~(\ref{eq:wfn}). The special case of $K^{(ho)}(\epsilon, l)=0$
for all energies corresponds to a pure harmonic oscillator.
\label{Figure1}}
\end{figure}

\begin{figure}
\scalebox{0.4}{\includegraphics{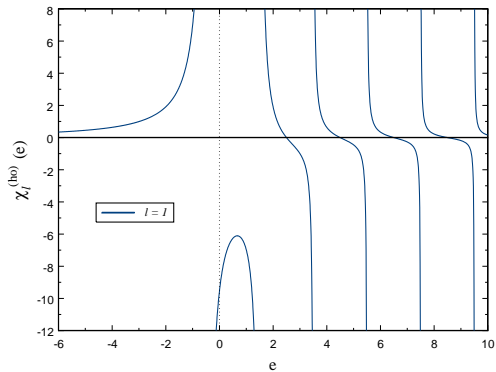}}
\caption{The same as Figure~1 except that it is for the $p$ wave.
\label{Figure2}}
\end{figure}

Making use of the large $r$ asymptotic behaviors of $f^{(ho)}$ 
and $g^{(ho)}$, as given by \cite{abr64}
\begin{eqnarray}
f^{(ho)}_{e l} &\stackrel{r\rightarrow\infty}{\longrightarrow}& 
	\frac{\Gamma(c)}{\Gamma(b)}x^{-e-1/2}e^{+x^2/2}  \;, 
\label{eq:fholr} \\
g^{(ho)}_{e l} &\stackrel{r\rightarrow\infty}{\longrightarrow}& 
	-\frac{2}{(2l+1)\pi}\left[
	\frac{\Gamma(2-c)}{\Gamma(1+b-c)}x^{-e-1/2}e^{+x^2/2}\right.\nonumber\\
	&-&\left.\frac{1}{\pi}\sin(\pi c)\Gamma(b)\Gamma(2-c)
	x^{+e-1/2}e^{-x^2/2}\right] \;,
\label{eq:gholr}	
\end{eqnarray}
and enforcing the boundary condition, $u_{\epsilon l}(r)\rightarrow 0$,
at infinity, we obtain the following equation that gives the
the energy spectrum of Eq.~(\ref{eq:rsch}) as the crossings
points of two functions:
\begin{equation}
\chi^{(ho)}_l(e) = K^{(ho)}(\epsilon,l) \;.
\label{eq:bsp}
\end{equation}
Here
\begin{equation}
\chi^{(ho)}_l(e) = (-1)^{l+1}[\Gamma(l+3/2)]^2
	\frac{\Gamma[1-(e+l+3/2)/2]}{\Gamma[(-e+l+3/2)/2]} \;,
\label{eq:chiho}
\end{equation}
is a universal function of the scaled energy $e$ that is determined
solely by the harmonic potential, and
$K^{(ho)}$ is the $K$ matrix that encapsulates all the short-range physics.
Plots of $\chi^{(ho)}_l(e)$ for the $s$ and $p$ waves are
shown in Figs.~\ref{Figure1} and \ref{Figure2}, respectively.

Equations~(\ref{eq:bsp}) and (\ref{eq:chiho}) give a rigorous formulation 
of energy spectrum for any potential that is asymptotically a harmonic 
oscillator.
It can be implemented numerically, or used as a basis for approximate
analytic solutions. The case of a pure symmetric harmonic oscillator 
is included as a special case,
corresponding to $K^{(ho)}(\epsilon,l)=0$ for all $\epsilon$,
with a well-known energy spectrum of
$\epsilon = \hbar\omega(2j+l+3/2)$, where $j=0,1,2,\dots$.

From Eq.~(\ref{eq:bsp}), it is clear that the key to finding the energy
spectrum is to find the $K$ matrix $K^{(ho)}$, as $\chi^{(ho)}_l$ is 
already known analytically. For this purpose, we note that 
the reference functions $f^{(ho)}$ and $g^{(ho)}$ have the
following small $r$ asymptotic behaviors that can be derived from a proper
expansion of the confluent hypergeometric function \cite{abr64}.
For $\epsilon>0$, we have
\begin{eqnarray}
f^{(ho)}_{e l} &\stackrel{r\ll\beta_{ho}}{\longrightarrow}& \sqrt{\frac{2}{\pi}}\:
	\frac{2^{l+1/2}\Gamma(l+3/2)}{(k\beta_{ho})^{l+1}}
	(k r)j_l(k r)  \;, 
\label{eq:fhopesr} \\
g^{(ho)}_{e l} &\stackrel{r\ll\beta_{ho}}{\longrightarrow}& \sqrt{\frac{2}{\pi}}\: 
	\frac{(k\beta_{ho})^{l}}{2^{l+1/2}\Gamma(l+3/2)}
	(k r)y_l(k r)  \;,
\end{eqnarray}
where $k = (2\mu\epsilon/\hbar^2)^{1/2}$.
For $\epsilon<0$, we have
\begin{eqnarray}
f^{(ho)}_{e l} &\stackrel{r\ll\beta_{ho}}{\longrightarrow}& 
	\frac{2^{l+1/2}\Gamma(l+3/2)}{(\kappa\beta_{ho})^{l+1}}
	(\kappa r)^{1/2}I_{l+1/2}(\kappa r)  \;, \\
g^{(ho)}_{e l} &\stackrel{r\ll\beta_{ho}}{\longrightarrow}& 
	\frac{(-1)^{l+1}(\kappa\beta_{ho})^{l}}{2^{l+1/2}\Gamma(l+3/2)}
	(\kappa r)^{1/2}I_{-l-1/2}(\kappa r)  \;,
\label{eq:ghonesr}	
\end{eqnarray}
where $\kappa = (-2\mu\epsilon/\hbar^2)^{1/2}$.
For interactions $V(r)$ that deviate from the harmonic
potential only in the region of $r\ll\beta_{ho}$, these
behaviors greatly facilitate the matching to the short-range
solution, from which $K^{(ho)}$ can be determined.
This is illustrated in our treatment of two identical atoms in a symmetric 
harmonic trap, to be presented in the next section.

\section{Two identical atoms in a symmetric harmonic trap \label{sec:tasht}}

Two interacting atoms in symmetric harmonic trap are described by
a Hamiltonian
\begin{eqnarray}
H &=& -\frac{\hbar^2}{2m_1}\nabla_1^2-\frac{\hbar^2}{2m_2}\nabla_2^2
	+\frac{1}{2}m_1\omega_1^2 r_1^2
	+\frac{1}{2}m_2\omega_2^2 r_2^2 \nonumber\\
  & &+v(|\mathbf{r}_1-\mathbf{r}_2|) \;,
\end{eqnarray}
where $v(r)$ represents the interaction between them.
To be specific, we restrict ourselves here to a class of problems for which
the atomic interaction is characterized, at large distances, 
by an attractive $1/r^6$ van der Waals potential
\begin{equation}
v(r)\stackrel{r\rightarrow\infty}{\longrightarrow} 
	-C_6/r^6 ,
\label{eq:vdW6}	
\end{equation}
which has an associated length scale of 
$\beta_6 = (2\mu C_6/\hbar^2)^{1/4}$ \cite{gao98a}.

For two atoms having the same trapping frequencies, namely 
$\omega_1=\omega_2=\omega$, which include of course the case of
two identical atoms of interest here, the center-of-mass motion and the 
relative motion are separable, and the solution of two identical atoms 
in a symmetric harmonic trap reduces to the solution of 
Eq.~(\ref{eq:rsch}) with
\begin{eqnarray}
V(r) &=& v(r)+\frac{1}{2}\mu\omega^2r^2 \nonumber\\
	&=& -C_6/r^6+\frac{1}{2}\mu\omega^2r^2\;,\; r\ge r_0 \;,
\label{eq:twoapot}
\end{eqnarray}
where $r_0$ represents the radius inside which the interactions
of shorter range than $\beta_6$, such as the $-C_8/r^8$ correction,
would come into play.

Since the $V(r)$ characterized by Eq.~(\ref{eq:twoapot}) 
is asymptotically a harmonic oscillator, it is
amenable for the QDT treatment of Sec.~\ref{sec:qdt}. 
In particular, the solution for the energy spectrum reduces to
finding the $K$ matrix $K^{(ho)}$ for the class of problems 
defined by Eqs.~(\ref{eq:vdW6}) and (\ref{eq:twoapot}).
This will be accomplished here by taking advantage of
the disparate length scales in the system.

For $r\ge r_0$, the potential $V(r)$ as given by Eq.~(\ref{eq:twoapot})
has two length scales. In addition to $\beta_6$ that is associated with
the van der Waals interaction, the harmonic trapping potential has a
length scale that can be taken either as the $\beta_{ho}$ defined earlier,
or as $a_{ho}=(\hbar/m\omega)^{1/2}=\beta_{ho}/\sqrt{2}$.
We will use both interchangeably, but will emphasize $a_{ho}$ 
for the sake of easier comparison with other results. 

For atoms in a typical magnetic or optical trap, $a_{ho}$ is of the
order of a micron, which is much greater than $\beta_6$ that is of 
the order of 100 au or about 50 nm. 
Under this condition of $\beta_6\ll a_{ho}$, which we will
call the limit of weak confinement, the van der Waals and the
trapping potentials operate on distinctive lengths scales.
In the region of $r\gg \beta_6$, the van der Waals interaction 
is negligible, and we have 
\begin{equation}
u_{II}(r) = A[f^{(ho)}_{e l}- K^{(ho)}(\epsilon,l) g^{(ho)}_{e l}]\;.
\label{eq:uII}
\end{equation}
In the region of $r_0\le r\ll a_{ho}$, 
the harmonic potential is negligible, and the interaction is
dominated by the van der Waals interaction. Here the wave
function can be written as \cite{gao01}
\begin{equation}
u_{I}(r) = B[f^{c(6)}_{\epsilon_s l}- K^{c}(\epsilon,l) g^{c(6)}_{\epsilon_s l}]\;.
\label{eq:uI}
\end{equation}
where $f^{c(6)}_{\epsilon_s l}$ and $g^{c(6)}_{\epsilon_s l}$ are the
reference functions for the van der Waals potential, 
and $K^c$ is the corresponding short-range $K$ matrix
\cite{gao98a,gao01,gao04a}.

For weak confinement defined by $\beta_6\ll a_{ho}$,
there exists an intermediate region $\beta_6 \ll r \ll a_{ho}$ 
in which either, or both, the van der Waals potential and the trapping 
potential can be ignored. In this region, the wave function can
be written either in the form of the inner solution as given
by Eq.~(\ref{eq:uI}), or in the form of the outer solution
as given by Eq.~(\ref{eq:uII}), and they must agree with each other.

Since $r/a_{ho}\ll 1$ in the intermediate region,
$f^{(ho)}_{e l}$ and $g^{(ho)}_{e l}$ are given by 
Eqs.~(\ref{eq:fhopesr})-(\ref{eq:ghonesr}).
In the same region, $r/\beta_6\gg 1$ and the reference functions
for the van der Waals potential are given for $\epsilon\ge 0$ by
\cite{gao98a,gaoupb}
\begin{eqnarray}
f^{c(6)}_{\epsilon_s l} &=& 
	\sqrt{\frac{2}{\pi k\beta_6}}
	(k r)[Z^{c(6)}_{ff}j_l(kr)+Z^{c(6)}_{fg}y_l(kr)]\;,
\label{eq:fcpelr} \\
g^{c(6)}_{\epsilon_s l} &=& 
	\sqrt{\frac{2}{\pi k\beta_6}}
	(k r)[Z^{c(6)}_{gf}j_l(kr)+Z^{c(6)}_{gg}y_l(kr)]\;,
\end{eqnarray}
and for $\epsilon< 0$ by \cite{gao98a,gaoupb}
\begin{eqnarray}
f^{c(6)}_{\epsilon_s l} &=& 
	\frac{1}{2}[W^{c(6)}_{f-}+2(-1)^l W^{c(6)}_{f+}]
	(r/\beta_6)^{1/2}I_{l+1/2}(\kappa r)\nonumber \\
	&+&\frac{1}{2}[W^{c(6)}_{f-}-2(-1)^l W^{c(6)}_{f+}]
	(r/\beta_6)^{1/2}I_{-l-1/2}(\kappa r) \;,\\
g^{c(6)}_{\epsilon_s l} &=& 
	\frac{1}{2}[W^{c(6)}_{g-}+2(-1)^l W^{c(6)}_{g+}]
	(r/\beta_6)^{1/2}I_{l+1/2}(\kappa r)\nonumber \\
	&+&\frac{1}{2}[W^{c(6)}_{g-}-2(-1)^l W^{c(6)}_{g+}]
	(r/\beta_6)^{1/2}I_{-l-1/2}(\kappa r) \;.
\label{eq:gcnelr}	
\end{eqnarray}
Here the $Z^{c(6)}$ and $W^{c(6)}$ matrices 
describe the propagation of a wave function in a $-C_6/r^6$ 
type of potential from small to large distances, 
and vice versa \cite{gao98a,gao00,gao01}.
Their elements of are all universal functions of a scaled energy 
$\epsilon_s=\epsilon/s_E$, where $s_E=(\hbar^2/2\mu)(1/\beta_6)^2$ 
is the energy scale associated with the van der Waals interaction.
Explicit expressions for the elements of the $Z^{c(6)}$
can be found in Ref.~\cite{gao00}.
The $W^{c(6)}$ matrix, which is related to the $W$ matrix
defined in Ref.~\cite{gao98a} by a linear transformation \cite{gao05a},
is given by
\begin{widetext}
\begin{eqnarray}
W^{c(6)}_{f-}(\epsilon_s) &=& \frac{2^{-1/2}G_{\epsilon l}(\nu)}
	{(X_{\epsilon l}^2+Y_{\epsilon l}^2)\sin\pi\nu}
	\left[ (1+M_{\epsilon l})
	\sin(\pi\nu/2)X_{\epsilon l}+(1-M_{\epsilon l})
	\cos(\pi\nu/2)Y_{\epsilon l}\right] , \\
W^{c(6)}_{f+}(\epsilon_s) &=& \frac{2^{-1/2}G_{\epsilon l}(\nu)\cos\pi\nu}
	{(X_{\epsilon l}^2+Y_{\epsilon l}^2)}
	\left[ (1-M_{\epsilon l})
	\sin(\pi\nu/2)X_{\epsilon l}+(1+M_{\epsilon l})
	\cos(\pi\nu/2)Y_{\epsilon l}\right] , \\
W^{c(6)}_{g-}(\epsilon_s) &=& \frac{2^{-1/2}G_{\epsilon l}(\nu)}
	{(X_{\epsilon l}^2+Y_{\epsilon l}^2)\sin\pi\nu}
	\left[ (1-M_{\epsilon l})
	\cos(\pi\nu/2)X_{\epsilon l}-(1+M_{\epsilon l})
	\sin(\pi\nu/2)Y_{\epsilon l}\right] , \\
W^{c(6)}_{g+}(\epsilon_s) &=& \frac{2^{-1/2}G_{\epsilon l}(\nu)\cos\pi\nu}
	{(X_{\epsilon l}^2+Y_{\epsilon l}^2)}
	\left[ (1+M_{\epsilon l})
	\cos(\pi\nu/2)X_{\epsilon l}-(1-M_{\epsilon l})
	\sin(\pi\nu/2)Y_{\epsilon l}\right] \;.
\end{eqnarray}
\end{widetext}
Here $M_{\epsilon l}=G_{\epsilon l}(-\nu)/G_{\epsilon l}(\nu)$,
with $\nu$, $X_{\epsilon l}$, $Y_{\epsilon l}$, and 
$G_{\epsilon l}$, all of which are functions of the scaled energy $\epsilon_s$,
being defined in Ref.~\cite{gao98a}.

Comparing, in the intermediate region, 
the inner solution given by Eqs.~(\ref{eq:uI}) and 
(\ref{eq:fcpelr})-(\ref{eq:gcnelr}) with the outer solution
given by Eqs.~(\ref{eq:uII}) and (\ref{eq:fhopesr})-(\ref{eq:ghonesr}),
we obtain, for $\epsilon>0$, 
\begin{equation}
K^{(ho)}= [\Gamma(l+3/2)]^2\frac{\tan\delta_l}{(e/2)^{l+1/2}} \;,
\label{eq:Khope}
\end{equation}
where
\begin{equation}
\tan\delta_l(\epsilon) = -( Z^{c(6)}_{fg}-Z^{c(6)}_{gg}K^{c} )
	(Z^{c(6)}_{ff} - Z^{c(6)}_{gf}K^{c})^{-1} \;,
\label{eq:Kphyo}
\end{equation}
is the physical K matrix for atomic scattering in free space 
as given in AQDT \cite{gao01}.	
For $\epsilon<0$, we obtain
\begin{widetext}
\begin{equation}
K^{(ho)}= (-1)^l\frac{[\Gamma(l+3/2)]^2}{(|e|/2)^{l+1/2}}
	\left\{\frac{\chi^{c(6)}_l-K^c-2(-1)^l[(W^{c(6)}_{f+}/W^{c(6)}_{g-})
		-K^c(W^{c(6)}_{g+}/W^{c(6)}_{g-})]}
	{\chi^{c(6)}_l-K^c+2(-1)^l[(W^{c(6)}_{f+}/W^{c(6)}_{g-})
		-K^c(W^{c(6)}_{g+}/W^{c(6)}_{g-})]}\right\} \;,
\label{eq:Khone}	
\end{equation}
\end{widetext}
where $\chi^{c(6)}_l(\epsilon_s)=W^{c(6)}_{f-}/W^{c(6)}_{g-}$ 
is the $\chi$-function that determines the molecular spectrum in the absence of
trapping \cite{gao01}.

Equations~(\ref{eq:bsp})-(\ref{eq:chiho}) and (\ref{eq:Khope})-(\ref{eq:Khone})
give a complete description of the energy spectrum for two idential atoms 
in a symmetric harmonic trap, 
from deeply-bound molecular states to highly excited trap states.
The only assumption in the theory is the assumption of weak confinement
as specified by $\beta_6/a_{ho}\ll 1$, which is well satisfied 
under all existing experimental conditions.

Other than the two energy scaling parameters $\hbar\omega$ and
$s_E$, the only parameter in the theory is $K^c(\epsilon,l)$,
which characterizes the interactions of shorter range than
$\beta_6$. It can be replaced by other equivalent
short-range parameters such as the quantum-defect 
$\mu^c(\epsilon,l)$ \cite{gao04b} or the 
$K^0_l(\epsilon,l)$ parameter \cite{gao98b,gao04c}.

\section{Results and discussions \label{sec:results}}

\subsection{Universal spectrum at the length scale of $\beta_6$}

The results given by Eqs.~(\ref{eq:bsp}), (\ref{eq:chiho}), 
(\ref{eq:Khope})-(\ref{eq:Khone}) can be written in
different forms that are convenient for different purposes.
For conceptual understanding, it is best to rearrange them
so that the entire energy spectrum is given by the solutions 
of a single equation
\begin{equation}
\chi^{(ho,6)}_l(e,\beta_6/a_{ho}) = K^c(\epsilon,l) \;,
\label{eq:usp}
\end{equation}
where
\begin{equation}
\chi^{(ho,6)}_l = 
	\frac{Z^{c(6)}_{fg}(\epsilon_s)-Z^{c(6)}_{ff}(\epsilon_s)\xi^{(ho)}_l(e)}
	{Z^{c(6)}_{gg}(\epsilon_s)-Z^{c(6)}_{gf}(\epsilon_s)\xi^{(ho)}_l(e)}\;,
\end{equation}
for $e>0$,
\begin{equation}
\chi^{(ho,6)}_l = 
	\frac{\chi^{c(6)}_l(\epsilon_s)-\alpha^{(ho)}_l(e)
	[W^{c(6)}_{f+}(\epsilon_s)/W^{c(6)}_{g-}(\epsilon_s)]}
	{1-\alpha^{(ho)}_l(e)[W^{c(6)}_{g+}(\epsilon_s)/W^{c(6)}_{g-}(\epsilon_s)]}\;,
\end{equation}
for $e<0$, and we have defined
\begin{equation}
\xi^{(ho)}_l(e) = (-1)^{l}(e/2)^{l}(|e|/2)^{1/2}
	\frac{\Gamma[1-(e+l+3/2)/2]}{\Gamma[(-e+l+3/2)/2]} \;,
\label{eq:xiho}
\end{equation}
and
\begin{equation}
\alpha^{(ho)}_l(e) = 2(-1)^l\frac{1-\xi^{(ho)}_l(e)}{1+\xi^{(ho)}_l(e)} \;.
\label{eq:alpha}
\end{equation}
The $\chi^{(ho,6)}_l$ function is a universal function that is 
applicable to any two identical atoms in a symmetric harmonic trap, 
provided they interact via the $-C_6/r^6$ type of van der Waals 
potential at large interatomic separations.
The strengths of interactions, as characterized by $C_6$
and $\omega$, play a role only through energy scaling parameters
$s_E$ and $\hbar\omega$. Specifically, the $\chi^{(ho,6)}_l$ 
function is made up terms that depend on energy 
through two different scaled energies, $\epsilon_s=\epsilon/s_E$ 
and $e=\epsilon/\hbar\omega$ that are related by 
$\epsilon_s = (\beta_6/a_{ho})^2 e$.
In other words, it is made up functions that vary on 
two distinctive energy scales:
the $\xi^{(ho)}$ and the $\alpha^{(ho)}$ functions that varies 
on the scale of $\hbar\omega$,
and the $Z^{c(6)}$ and $W^{c(6)}$ matrix elements that vary on a
scale of $s_E$.  

The solutions, more precisely the inverse, of Eq.~(\ref{eq:usp}) can be written as
\begin{equation}
e = \widetilde{\Omega}^{(l)}_i(K^c,\beta_6/a_{ho}) \;,
\label{eq:uspec}
\end{equation}
where, similar to $\chi^{(ho,6)}_l$, the $\widetilde{\Omega}^{(l)}_i$ 
are universal functions that are uniquely determined by
the exponent of the van der Waals interaction 
($n=6$), and the exponent of the trapping potential (2 for the harmonic trap).
Instead of the parameter $K^c$, the same universal functions can also be expressed in
terms of equivalent parameters such as the quantum-defect $\mu^c$ \cite{gao04b}, 
or the $K^0_l$ parameter \cite{gao98b}, all of which are well defined for all $l$.

When the energy and angular momentum dependence of the short-range parameter
$K^c$ is included, Eqs.~(\ref{eq:usp}) and (\ref{eq:uspec}) are exact, 
and applicable to arbitrary energy and partial waves.
Ignoring the energy and the $l$ dependence of $K^c$,
which in the case of a single channel is due entirely to interactions
of shorter range than $\beta_6$ \cite{gao01,gao04b}, 
the solutions of Eq.~(\ref{eq:usp}),
namely Eq.~(\ref{eq:uspec}) with $K^c=K^c(\epsilon=0,l=0)$, 
or its variations, give what we
call the universal spectrum at length scale $\beta_6$ \cite{gao01,gao04a,kha06}.
It is followed by all two-atom systems in a trap with $-C_6/r^6$ type
of long-range interaction, over a range of energies that is hundreds
of $s_E$ around the threshold \cite{gao01,gao05a},
which far exceeds all energies of interest in cold-atom physics.
(As an example, $s_E = 9.331\times 10^{-4}$ K for $^{23}$Na.)
Other than the energy scaling parameter $s_E$ that is determined
by the $C_6$ coefficient and the atomic mass,
all energy levels in this energy range, including states of different
$l$, are determined, in the case of single channel, by two parameters,
$K^c=K^c(\epsilon=0,l=0)$ and $\beta_6/a_{ho}$.
This is an example of universal spectrum
at the second longest length scale in the system \cite{kha06}, 
as opposed to the universal spectrum at the longest 
length scale, which would have required only a single parameter \cite{gao01}.
 
Depending on the physics of interest, the universal spectrum at length
scale $\beta_6$ can also be expressed in terms of other parameters.
In particular, it can be expressed in terms of the $s$ wave 
scattering length, $a_0$, as
\begin{equation}
e = \Omega^{(l)}_i(a_{0}/a_{ho},\beta_6/a_{ho}) \;,
\label{eq:usps}
\end{equation}
since $a_0$ can be related to $K^c=K^c(\epsilon=0,l=0)$ by
\begin{equation}
a_{0}/\beta_n = \left[b^{2b}\frac{\Gamma(1-b)}{\Gamma(1+b)}\right]
	\frac{K^c(0,0) + \tan(\pi b/2)}{K^c(0,0) - \tan(\pi b/2)} \;,
\label{eq:a0sKc}
\end{equation}
where $b=1/(n-2)$ with $n=6$ \cite{gao03a,gao04b}.
Similar representations of universal spectrum can also be defined
more generally for $N$ atoms ($N>2$) in a trap \cite{gao04a,gao05b,kha06}.

\begin{figure}
\scalebox{0.4}{\includegraphics{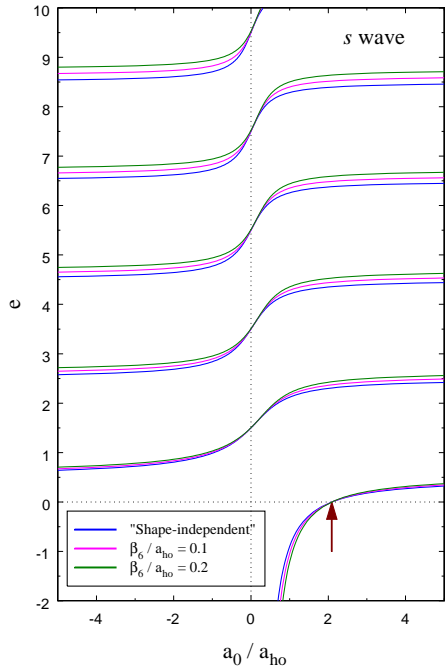}}
\caption{Universal $s$ wave spectrum at length scale $\beta_6$ for 
	two identical atoms in a symmetric harmonic trap and with an asymptotic 
	interaction of the type of $-1/r^6$. The arrow points to the
	$s$ wave scattering length, $a_{0x}$, beyond which the least-bound 
	molecular $s$ state is pushed to positive energies.
\label{Figure3}}
\end{figure}

Figure~\ref{Figure3} illustrates the 
universal $s$ wave spectrum at length scale $\beta_6$ for 
two identical atoms in a symmetric harmonic trap. 
It uses the representation of Eq.~(\ref{eq:usps}) to facilitate
comparison with the shape-independent approximation 
(see Ref.~\cite{bus98} and Sec.~\ref{sec:sia}).
The $ \Omega^{(l=0)}$ functions of two variables 
$a_{0}/a_{ho}$ and $\beta_6/a_{ho}$ are plotted here as functions 
of $a_{0}/a_{ho}$ for different values of $\beta_6/a_{ho}$.
In the small range of energies shown in the figure, which
corresponds to $\epsilon\sim \hbar\omega \ll s_E$, 
the universal spectrum
approaches that of the shape-independent results \cite{bus98}
in the limit of $\beta_6/a_{ho}\rightarrow 0$ \cite{kha06}. 

\begin{figure}
\scalebox{0.4}{\includegraphics{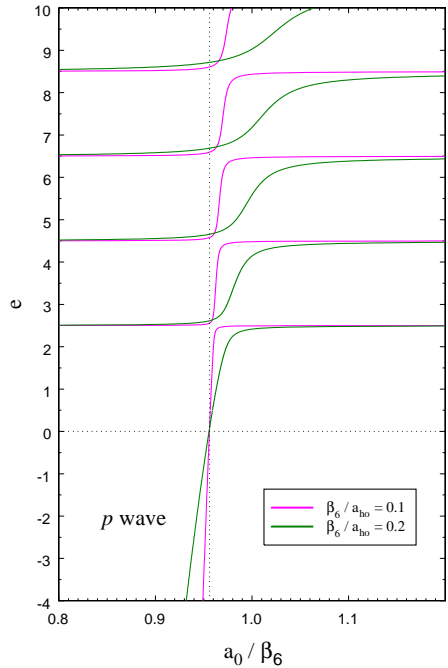}}
\caption{Universal $p$ wave spectrum at length scale $\beta_6$ for 
	two identical atoms in a symmetric harmonic trap and with an asymptotic 
	interaction of the type of $-1/r^6$, plotted here versus a scaled
	$s$ wave scattering length, $a_0/\beta_6$.
\label{Figure4}}
\end{figure}

\begin{figure}
\scalebox{0.4}{\includegraphics{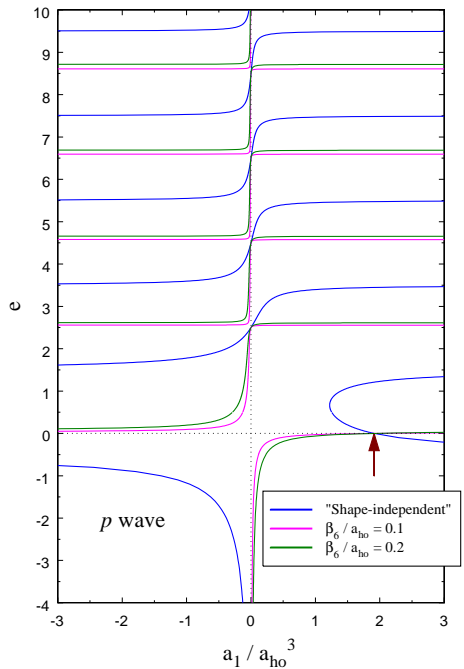}}
\caption{Universal $p$ wave spectrum at length scale $\beta_6$ for 
	two identical atoms in a symmetric harmonic trap and with an asymptotic 
	interaction of the type of $-1/r^6$, plotted here versus a
	scaled $p$ wave scattering length (volume) $a_1/a_{ho}^3$. 
	The arrow points to the
	$p$ wave scattering length, $a_{1x}$, beyond which the least-bound 
	molecular $p$ state is pushed to positive energies.
\label{Figure5}}
\end{figure}

Figures~\ref{Figure4} and \ref{Figure5} both illustrate the 
universal $p$ wave spectrum at length scale $\beta_6$. 
They also serve to illustrate how
different representations of the universal spectrum can serve
different purposes in terms of physical understanding.
In Fig.~\ref{Figure4}, the universal $p$ wave spectrum is plotted
as a function of $a_0/\beta_6$ for different values of $\beta_6/a_{ho}$.
It gives the best illustration that, in the case of single channel,
the $s$ wave scattering length determines not only the $s$ wave
spectrum, but also the spectra of other partial waves 
including the $p$ wave \cite{gao01}. 
It also determines the $p$ wave scattering length $a_1$ 
(also called scattering volume
as it has the dimension of a volume) through \cite{gao04c}
\begin{equation}
a_{1}/\beta_6^3 = \frac{[\Gamma(1/4)]^2}{18\pi}
	\frac{a_{0}-\bar{a}_{0}}{2\bar{a}_{0}-a_{0}} \;,
\label{eq:a1a0}	
\end{equation}
where $\bar{a}_{0} = 2\pi\beta_6/[\Gamma(1/4)]^2 = 0.4779888 \beta_6$
is the mean $s$ wave scattering length of Gribakin and Flambaum \cite{gri93}.
Figure~\ref{Figure4} shows that the $p$ wave spectrum, 
in the case of single channel,
is strongly influenced by the atomic interaction only when the
$s$ wave scattering length is around $2\bar{a}_{0}=0.955978\beta_6$,
which corresponds to having a $p$ wave bound state right at the
threshold \cite{gao00,gao04b}.
For $a_0$ slightly larger than $2\bar{a}_{0}$, $a_1$ is large and 
negative [see Eq.~(\ref{eq:a1a0})],
the lowest few trap states are strongly affected by a $p$ wave shape
resonance near the threshold \cite{gao98b}.
For $a_0$ slightly less than $2\bar{a}_{0}$, $a_1$ is large and positive,
and there is a $p$ wave molecular state close to the threshold \cite{gao04c}.

In Fig.~\ref{Figure5}, the universal $p$ wave spectrum is plotted
as a function of $a_1/a_{ho}^3$ for different values of $\beta_6/a_{ho}$.
The advantages of this representation are twofold.
First, it facilitates comparison with the $p$ wave shape-independent
approximation to be discussed in Sec.~\ref{sec:sia}.
Second, while the representation shown in Fig.~\ref{Figure4}
is applicable only in the case of single channel,
the representation shown in Fig.~\ref{Figure5} 
would apply even in multichannel cases, provided the energy dependence 
of $K^c$ due to closed channels is negligible in energy range of 
interest \cite{gao05a,gaoupb}. This is because in representing
the $p$ wave spectrum as a function of $a_1$, instead of $a_0$, one
is only making use of the energy-independence of $K^c$,
not its angular-independence, which takes on different characteristics
(still related) for multichannel problems \cite{gao05a,gaoupb}.
Specifically, this representation is obtain from solving Eq.~(\ref{eq:usp})
using $K^c=K^c(\epsilon=0,l=1)$, which is related to the $a_1$ through
the following relations \cite{gao98b,gao04c}
\begin{equation}
a_{1} = -\bar{a}_1\left(1+\frac{1}{K^0_{l=1}(\epsilon=0)}\right) \;,
\end{equation}
where $\bar{a}_1=[\Gamma(1/4)]^2\beta_6^3/(36\pi)$ is a mean $p$ wave
scattering length, and
\begin{equation}
K^0_l(\epsilon=0) = \frac{c_l-K^c(\epsilon=0,l)}{1+c_lK^c(\epsilon=0,l)}\;,
\label{eq:K0l6}
\end{equation}
where $c_l=\tan(\frac{1}{4}l\pi+\frac{1}{8}\pi)$.

Comparing either Fig.~\ref{Figure4} or \ref{Figure5} with
Fig.~\ref{Figure3} leads to one of the more important conclusions
of this work. That is, unlike the noninteracting particles in a trap for
which the lowest $p$ state energy is always greater than that of
the $s$ states, a strong $p$ wave pairing ($a_1\sim a_{ho}^3$ or greater) 
for interacting particles can 
lead to a lower energy state around the threshold than 
a $s$ wave pairing. 
This state is, for large and negative $a_1$, a $p$ wave shape
resonance stabilized by the trap, and is a $p$ wave molecular state
for large and positive $a_1$.
Similar statements can also be made for other, higher partial waves.
Furthermore, we expect the same physics to persist in
many-atom systems, which will be a subject of future investigations.
Other characteristics of the universal spectra are addressed 
separately in subsequent sections.

\subsection{Limitations of ``shape-independent'' approximations}
\label{sec:sia}

All previous results on two atoms in a symmetric harmonic trap 
can be easily derived as various approximations within our theory.
In particular, the ``shape-independent'' approximation corresponds 
to ignoring energy dependence of $K^{(ho)}$, and taking it to be 
its value at zero energy, namely,
\begin{equation}
K^{(ho)}\approx -\frac{[\Gamma(l+3/2)]^2}{(\beta_{ho}/2)^{2l+1}}
	\lim_{k\rightarrow 0}\left[-\frac{\tan\delta_l}{k^{2l+1}}\right] \;.
\label{eq:Khosia}
\end{equation}
For partial waves with well-defined scattering lengths,
$a_l =\lim_{k\rightarrow 0}[-\tan\delta_l/k^{2l+1}]$, 
this approximation used in Eq.~(\ref{eq:bsp}) leads to the following
equation for the energy spectrum:
\begin{equation}
(-1)^{l}\left(\frac{1}{2}\right)^{l+1/2}\frac{\Gamma[1-(e+l+3/2)/2]}{\Gamma[(-e+l+3/2)/2]} 
	= a_l/a_{ho}^{2l+1} \;,
\label{eq:sia}
\end{equation}
for both positive and negative energies.
For $l=0$, Eq.~(\ref{eq:sia}) reproduces the result of Ref.~\cite{bus98},
derived using a delta-function pseudopotential \cite{hua57}.

Figure~\ref{Figure3} has shown that for the $s$ wave, the shape-independent
approximation gives a good approximation to the universal spectrum
under weak confinement ($\beta_6/a_{ho}\ll 1$) and for energies
$|\epsilon|\sim \hbar\omega\ll s_E$.
The shape dependence is more important for strong coupling
($a_0\sim a_{ho}$ or greater), 
but especially for energies further away from the threshold.
\begin{figure}
\scalebox{0.4}{\includegraphics{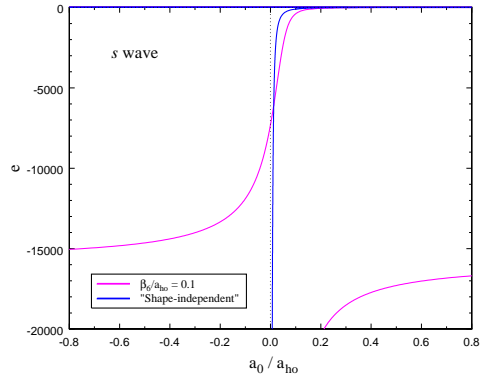}}
\caption{Universal $s$ state spectrum at length scale $\beta_6$ for 
	two identical atoms in a symmetric harmonic trap and with an asymptotic 
	interaction of the type of $-1/r^6$, plotted to illustrate the failure
	of the shape-independent approximation away from the threshold.
\label{Figure6}}
\end{figure}
In particular, the shape-independent approximation breaks down for all energies
$\epsilon\sim s_E$ or greater, for which the energy dependence of
$K^{(ho)}$ due to the long-range van der Waals interaction can no
longer be ignored \cite{gao98b,gaoupb}. For example, 
it does not give the proper molecular binding energy for small 
positive scattering lengths, and it fails completely to describe 
molecular states of negative scattering length, as illustrated in 
Fig.~\ref{Figure6}.
In reality, these more deeply bound molecular states approach
those of a free molecule in the absence of a trap, with the trapping
potential serving as a weak perturbation, as to be discussed 
further in Sec.~\ref{sec:molspec}.

With the relative success of the ``shape-independent'' approximation 
for the $s$ wave in the threshold region, it is important
to emphasize its severe limitations for any partial waves
other than the $s$ wave. 
For atoms with $-C_6/r^6$ type of van der Waals interaction, 
the ``shape-independent'' approximation 
clearly fails for $l\ge 2$, for which there are no well-defined
scattering lengths \cite{lev63,gao98b}. 
Even for the $p$ wave, it is applicable only at zero
energy, or for weak $p$ wave coupling as characterized by 
$a_1\sim\beta_6^3\ll a_{ho}^3$ or smaller, 
as illustrated in Fig.~\ref{Figure4}.
This failure of the ``shape-independent'' approximation 
for $l>0$ is directly related to the failure of the effective range
theory (ERT) \cite{bla49} in describing the shape resonances close to
the threshold, and more generally to the failure of
ERT in describing Feshbach resonances in nonzero partial
waves \cite{gaoupb}.

Our theory allows for simple generalizations beyond the shape-independent
approximation. For example, for positive energies,
Eqs.~(\ref{eq:bsp}), (\ref{eq:chiho}) and (\ref{eq:Khope}) 
that determine the energy spectrum can be rewritten as
\begin{equation}
-\xi^{(ho)}_l(e) = \tan\delta_l(\epsilon) \;.
\label{eq:esppe}
\end{equation}
Instead of the effective-range expansion for $\tan\delta_l$,
which leads to the ``shape-independent'' approximation,
one can simply use the corresponding QDT expansion for
$-C_6/r^6$ type of potential \cite{gao98b}.
The results would be applicable from zero energy up to 
$\epsilon\sim s_E$ 
for the $s$ wave, and over a greater range of energies for higher
partial waves. This energy range, while much smaller than that
described by the universal spectrum Eq.~(\ref{eq:usp}), 
already exceeds the range of interest in existing experiments \cite{sto06}.

\subsection{Trap states and molecular states}

The states of two atoms in a trap can be classified into trap states
and molecular states. The former corresponds to states that
evolve into diatomic continuum as the trap is turned off 
adiabatically ($a_{ho}\rightarrow\infty$).
The latter corresponds to states that evolve into bound molecular
states in the same limit. Note that the molecular states would not
have existed in the hard-sphere atomic model \cite{blo02}.

The molecular state of highest energy correspond to the ones in
Figs.~\ref{Figure3}-\ref{Figure5} that cross the zero energy.
It is the least-bound molecular state that gets pushed up in energy
by the trapping potential.
The crossing into positive energy can happen either by
tuning up the scattering length, which has the effect of making
the binding energy of the molecule sufficiently small in the absence
of the trap, or by tuning up the trap frequency.

The crossing points, which correspond to having a state right at $\epsilon=0$,
can be easily found through Eq.~(\ref{eq:sia}),
which is exact at zero energy for partial waves with well defined scattering
lengths.
For $s$ and $p$ waves, it gives
\begin{equation}
a_{lx} = (-1)^{l}\left(\frac{1}{2}\right)^{l+1/2}\frac{\Gamma(1/4-l/2)}{\Gamma(l/2+3/4)}a_{ho}^{2l+1} \;.
\end{equation}
For the $s$ wave with a fixed trap frequency, it gives
\begin{equation}
a_{0x} = \frac{[\Gamma(1/4)]^2}{2\pi}a_{ho} =  2.0920992\:a_{ho} \;,
\end{equation}
beyond which the least bound molecular states is pushed to positive
energy.
For a fixed $a_0\gg \beta_6$, the same equation determines the crossing
trap frequency
\begin{equation}
\omega_{0x} = \frac{[\Gamma(1/4)]^4}{(2\pi)^2}\frac{\hbar}{m a_0^2} \;,
\end{equation}
beyond which the least bound $s$ wave molecular state is pushed into positive
energy.

For the $p$ wave, it gives
\begin{equation}
a_{1x} = \frac{8\pi}{[\Gamma(1/4)]^2}a_{ho}^3
	=  1.9119552\:a_{ho}^3 \;.
\end{equation}
For a fixed $a_1\gg \beta_6^3$, it determines the crossing
trap frequency
\begin{equation}
\omega_{1x} = \frac{4\pi^{2/3}}{[\Gamma(1/4)]^{4/3}}\frac{\hbar}{m a_1^{2/3}} \;,
\end{equation}
beyond which the least bound $p$ wave molecular state is pushed to
positive energy.

All higher branches of states in Figs.~\ref{Figure3}-\ref{Figure5} 
are trap states. All lower branches, which approach those of
molecular states in the absence of trapping \cite{gao01}, as discussed in 
the next section, are molecular states.

\subsection{Trapping shift of molecular spectrum}
\label{sec:molspec}

Except for the least-bound state with binding energy comparable to 
or smaller than $\hbar\omega$, the effects of trapping on  
molecular states are generally weak, and can be treated 
perturbatively. The finite range of such states is such that the
atoms in them would hardly feel the existence of
a trap in their relative motion. 

The nature of this perturbation is best understood
by rewriting Eq.~(\ref{eq:usp}) for $e<0$ as
\begin{equation}
\chi^{c(6)}_l(\epsilon_s) = K^c
	+\alpha^{(ho)}_l(e)
	\frac{W^{c(6)}_{f+}-K^cW^{c(6)}_{g+}}{W^{c(6)}_{g-}} \;.
\label{eq:uspne}	
\end{equation}
Comparing this equation to $\chi^{c(6)}_l(\epsilon_s) = K^c$,
which determines the molecular spectrum in free space \cite{gao01},
the effect of trapping is isolated in this formulation
to the second term in Eq.~(\ref{eq:uspne}).
From Eq.~(\ref{eq:alpha}) and 
\begin{equation}
\xi^{(ho)}_l(e) \stackrel{-e\gg 1}{\longrightarrow}
	1+\frac{1}{6}\frac{(l-1/2)(l+1/2)(l+3/2)}{|e|^2}\;,
\end{equation}
it is clear, as expected, that the molecular states 
with binding much greater than $\hbar\omega$
are only weakly affected by the trapping, and can be
treated by solving Eq.~(\ref{eq:uspne}) perturbatively.
To the lowest order in $\beta_6/a_{ho}$, the energy shift 
due to trapping is given by 
\begin{equation}
\Delta\epsilon_{ls} = q_l(\epsilon_{ls})
	\left(\frac{\beta_6}{a_{ho}}\right)^4 \;,
\label{eq:de}	
\end{equation}
where
\begin{equation}
q_l(\epsilon_{ls}) = (-1)^l
	\frac{(l-1/2)(l+1/2)(l+3/2)}{6|\epsilon_{ls}|^2
	\left.\left[W^{c(6)}_{f-}W^{c(6)\prime}_{g-}
	-W^{c(6)}_{g-}W^{c(6)\prime}_{f-}\right]\right|_{\epsilon_s=\epsilon_{ls}}} \;,
\label{eq:eta}	
\end{equation}
and $\epsilon_{ls}=\epsilon_l/s_E$ is the scaled bound state energy of a 
molecule in the absence of the trap.
Equation~(\ref{eq:de}) means that trapping shift is, to the lowest order, 
proportional to $(\beta_6/a_{ho})^4$, multiplied by a universal function of
the scaled binding energy that goes to zero in the limit of 
$|\epsilon_{ls}|\rightarrow\infty$.

\subsection{Highly excited trap states}

For an highly excited trap state with $e\gg 1$,
the determination of energy spectrum can also be further simplified.
From 
\begin{equation}
\xi^{(ho)}_l(e)\stackrel{e\gg 1}{\longrightarrow} \tan[\pi(e-l+1/2)/2)],
\end{equation}
Eq.~(\ref{eq:esppe}) reduces, for $e\gg 1$, to
\begin{equation}
\tan[\pi(e-l+1/2)/2] = -\tan[\delta_l(\epsilon)]\;,
\end{equation}
or $(e-l+1/2)\pi/2+\delta_l(\epsilon) = j\pi$, where $j$ is an integer.
This result for $l=0$ has also been derived by Bolda \textit{et al.}
using a generalized pseudopotential \cite{bol02}.
It could also have been derived by using a semiclassical approximation for
the harmonic part of the potential.
For all energies of experimental interest, the phase shift
$\delta_l$ can be accurately described using AQDT parametrization 
of Ref.~\cite{gao01}.

\subsection{Long-range correlation between atoms in a trap}

In studies of quantum few-body and quantum many-body systems,
it is often assumed that the wave function can be written in
a Jastrow form \cite{jas55}, which is given, e.g., for bosons by
\begin{equation}
\Psi = \left[\prod_{i=1}^N \phi(\mathbf{r}_i)\right]
	\prod_{i<j=1}^NF(r_{ij}) \;.
\end{equation}
Here $N$ is the number of particles, $\phi$ represents an 
independent-particle orbital,
and $F$ is the pair correlation function.
It is commonly assumed that
$F$ has the asymptotic behavior of $F(r) \longrightarrow 1$ 
at large $r$, meaning that the particles become uncorrelated at large 
separations \cite{jas55}.

Our theory here provides a opportunity to check the validity of
these assumptions, at least for $N=2$. To be specific, we restrict 
ourselves here to the lowest $s$ wave trap state
for which the Jastrow assumptions are usually applied.
Combining the wave function for the relative motion, as given
by Eqs.~(\ref{eq:wfn})-(\ref{eq:gholr}),
and that for the center-of-mass motion, which is a pure harmonic
oscillator in its ground state, it is easy to show that the total
wave function for two identical atoms in a symmetric harmonic trap
can indeed be written in the Jastrow form but with
a correlation function that behaves as  
\begin{equation}
F(r) \longrightarrow Cr^{e-3/2} \;,
\end{equation}
in the limit of large interatomic separations. 
Since $e$ can deviate significantly from
3/2 for strong coupling ($a_0/a_{ho}\sim 1$ or greater), 
as illustrated in Fig.~\ref{Figure3}, 
this result implies that there 
exists significant long-range correlation for strongly-interacting
particles in a trap.
In a recent publication \cite{kha06}, we have shown that such long-range 
correlation exists not only for two particles in a trap,
but also for $N$ ($N>2$) 
strongly-interacting particles in a trap.

\section{Conclusions \label{sec:conclusions}}

In conclusion, we have presented a QDT for a harmonic
potential and a two-scale QDT for 
two identical atoms in a symmetric harmonic trap.
It is a general theory that is applicable to different partial waves,
and from deeply bound molecular states to highly excited trap states.
The only approximation in the theory, $\beta_6/a_{ho}\ll 1$,
can be relaxed if needed. The result will still be of the form of
Eq.~(\ref{eq:usp}) except that the $\chi^{(ho,6)}$ 
function for $\beta_6/a_{ho}\sim 1$ or greater will require
a more general solution of the two-scale potential described by
Eq.~(\ref{eq:twoapot}).
If the scattering length parameter used in our formulation is 
achieved by tuning around a Feshbach resonance \cite{stw76,tie93},
the same results would apply, provided it is a broad 
Feshbach resonance with a width much greater than the energy scale 
$s_E$ associated with the van der Waals potential \cite{koh04,sim05,julbg06}.

From a more general perspective, the theory we have presented here 
demonstrates the
concept of multiscale QDT, which has potential implications
for a range of problems. As examples, we mention 
two other two-scale potentials:
\begin{equation}
V(r) = -C_6/r^6-C_8/r^8 \;,
\end{equation}
and
\begin{equation}
V(r) = -Z/r-C_4/r^4 \;.
\end{equation}
The former is of interest for a more accurate QDT description
of long-range molecules over a wider range of energies, 
the latter is of interest for a more systematic understanding
of the core polarization effect on atomic spectra.
While such potentials have been well studied in numerical calculations,
a more systematic approach to such two-scale problems may well 
be worthy of future efforts. 

\begin{acknowledgments}
We thank Eite Tiesinga, Eric Bolda, and Paul Julienne for helpful discussions.
This work was supported by the National Science Foundation under 
the Grant No. PHY-0457060.
\end{acknowledgments}

\bibliography{sac,bose}

\end{document}